\documentclass[preprintnumbers,twocolumn,prd]{revtex4}
\usepackage{amssymb,amsmath}
\usepackage[dvips]{graphicx,color}

\newcommand{\Lag}{\mathcal{L}}
\newcommand{\Ham}{\mathcal{H}}
\newcommand{\D}{\mathcal{D}}
\newcommand{\tr}{\mathrm{tr}}
\newcommand{\Pcal}{\mathcal{P}}
\newcommand{\Nc}{N_{\text{c}}}
\newcommand{\LQCD}{\Lambda_{\text{QCD}}}
\newcommand{\energy}{\varepsilon}

\newcommand{\Vad}{{V_{\text{A}}}}
\newcommand{\tf}{T_{\text{F}}}

\newcommand{\xt}{\boldsymbol{x}_\perp}
\newcommand{\yt}{\boldsymbol{y}_\perp}

\newcommand{\qt}[1]{\boldsymbol{q}_{#1\perp}}
\newcommand{\kt}{\boldsymbol{k}_\perp}
\newcommand{\del}{\boldsymbol{\partial}}
\newcommand{\zero}{\boldsymbol{0}_\perp}

\newcommand{\MeV}{\;\text{MeV}}
\newcommand{\GeV}{\;\text{GeV}}
\newcommand{\fm}{\;\text{fm}}

\newcommand{\half}{{\tfrac{1}{2}}}
\newcommand{\quart}{{\tfrac{1}{4}}}
\begin{document}

\title{Initial energy density and gluon distribution from the Glasma
  in heavy-ion collisions }
\author{Hirotsugu Fujii}
\affiliation{Institute of Physics, University of Tokyo,
             Komaba, Meguro-ku, Tokyo 153-8902, Japan}
\author{Kenji Fukushima}
\affiliation{Yukawa Institute for Theoretical Physics,
             Kyoto University, Kyoto 606-8502, Japan}
\author{Yoshimasa Hidaka}
\affiliation{RIKEN BNL Research Center,
             Brookhaven National Laboratory, Upton, NY 11973, USA}
\begin{abstract}
 We estimate the energy density and the gluon distribution associated
 with the classical fields describing the early-time dynamics of the
 heavy-ion collisions.  We first decompose the energy density into the
 momentum components exactly in the McLerran-Venugopalan model, with
 the use of the Wilson line correlators.  Then we evolve the energy
 density with the free-field equation, which is justified by the
 dominance of the ultraviolet modes near the collision point.  We also
 discuss the improvement with inclusion of nonlinear terms into the
 time evolution.  Our numerical results at RHIC energy are fairly
 consistent with the empirical values.
\end{abstract}
\preprint{YITP-08-51, UT-Komaba/08-17, RBRC-761}
\maketitle

%%%%%%%%%%   INTRODUCTION   %%%%%%%%%%

\section{INTRODUCTION}

Color Glass Condensate (CGC) provides us with a theoretical foundation
for the weak-coupling description of soft partons associated with
highly energetic hadrons~\cite{review,McLerran:1993ni}.  The number of
the soft partons grows with increasing energy due to quantum
branching, which eventually leads to merging and saturation of the
partons in the projectile at a certain energy characterized by the
scale $Q_s$.  Such soft (wee) partons are relevant not only in the
diffractive process in deep inelastic scatterings but also for
thermalization in relativistic heavy-ion
collisions~\cite{Gyulassy:2004zy}.  In the latter case, the dense
transient system at mid-rapidity is initially created through the
interactions between the soft partons from the incident nuclei.

In the Au-Au collisions at the top energy $\sqrt{s_{_{NN}}}=200\GeV$
of Relativistic Heavy Ion Collider (RHIC), the relevant scale of
Bjorken's $x$ is roughly estimated as
$p_\perp/\sqrt{s_{_{NN}}}\sim 10^{-2}$ with presuming that the bulk of
the initial medium is composed of the gluons of momenta
$p_\perp\lesssim1\GeV$.  Then, the phenomenological
Golec-Biernat--W\"{u}sthoff
fit~\cite{GolecBiernat:1998js,Stasto:2000er}, multiplied by a nuclear
enhancement factor $A^{1/3}=5.8$ (see
Refs.~\cite{Freund:2002ux,Kowalski:2007rw}) gives rise to
$Q_s^2\simeq 2\GeV^2$.  This implies that the bunch of the incoming
gluons with $p_\perp\lesssim Q_s$ in the incident nuclei are in the
saturation regime.  These gluons are to be described as the classical
Weizs\"{a}cker-Williams fields~\cite{Kovner:1995ja,Gyulassy:1997vt} in
the first approximation.  Such a classical field picture is expected
to be more reliable at the energy of Large Hadron Collider (LHC),
$\sqrt{s_{_{NN}}}=5500\GeV$.  At this energy the saturation scale for
the Pb-Pb collisions will be around $Q_s^2\simeq 5.2\GeV^2$, i.e.,
$(1.6)^2$ times larger than that of RHIC.\ \ Since it will be shown
that the energy density and multiplicity will scale with $Q_s^3$ and
$Q_s^2$, respectively, the former will become $1.6^3=4.1$ times larger
and the latter $1.6^2=2.6$ times larger at LHC than at RHIC.\ \ At the
same time, the typical time scale at LHC will be $1.6^{-1}=0.63$ times
shorter than at RHIC~\cite{Kharzeev:2004if,Lappi:2008eq}.  In this
paper we will address the initial stage of the nuclear collisions at
RHIC energy taking the saturation scale as $Q_s^2=1$--$2\GeV^2$.
[This uncertainty comes from the choice of relevant $x$.]

There are a number of efforts to solve the classical Yang-Mills
equations of motion in order to determine the early-time evolution in
the heavy-ion collisions~\cite{Kovner:1995ja,Gyulassy:1997vt}.  The
numerical studies have been performed quite successfully so
far~\cite{Krasnitz:1998ns,Krasnitz:2001qu,Lappi:2003bi,Lappi:2006fp,%
Lappi:2006hq} and also the analytical techniques are developing
recently~\cite{Fries:2006pv,Fukushima:2007ja,Fujii:2008dd} based on
the McLerran-Venugopalan (MV) model~\cite{McLerran:1993ni}
in the CGC picture.

This paper is a continued attempt from Ref.~\cite{Fukushima:2007ja} to
give an analytical estimate for the energy deposit and the gluon
production from the colliding coherent fields.  It is well known that
the initial gauge configuration contains only the longitudinal
fields~\cite{Kovner:1995ja,Gyulassy:1997vt,Lappi:2006fp,Fukushima:2006ax}.
These longitudinal fields rapidly decay and the nonzero transverse
fields are generated in the early-time evolution.  However, if we
expand the evolution of the energy density in a power series of $\tau$
from the collision time $\tau=0$, we find that an ultraviolet (UV)
divergence in each term, especially in the zeroth order term,
i.e.\ initial energy density.  This divergence originates from the
perturbative tail of the gluon distribution in the MV model.

In Ref.~\cite{Fukushima:2007ja} an ansatz of the logarithmic form was
proposed by one of the authors to resum this UV singularity.  Here we
take another way.  Because the origin of the divergence is in the UV
regime where couplings between the classical fields are unimportant,
one should be able to tame the divergent terms within the perturbative
framework.  This is indeed the case, as argued in
Ref.~\cite{Kovchegov:2005ss} in a different context and emphasized
also in the numerical simulation~\cite{Lappi:2006hq}.  In this paper
we shall resum all the most UV-singular terms, which results in
solving the free-field equation.  The solution is the Bessel function
in the boost-invariant expanding geometry.  At finite proper time
$\tau$ we can send the UV cutoff to infinity.  Moreover we find that
the energy density behaves as $1/\tau$ at large $\tau$, consistently
with the free-streaming behavior.  In fact, the limit $\tau\to 0$
reproduces the original UV divergence of the energy density as it
should.

Our present approach consists of the exact initial condition of the MV
model and the following perturbative time evolution.  The saturation
effect is fully taken into account in the initial condition in this
sense.  We first neglect any nonlinear effect in the time dependence,
and therefore the validity of this first treatment is limited to the
very early time when the UV modes dominate the dynamics.  This
approach recovers the same analytical result as
Ref.~\cite{Kovchegov:2005ss} in a simple manner.  Although the
free-streaming region is no longer in the reliable range of the
approximation, it is worth estimating the energy and the gluon
multiplicity there and comparing them with the empirical values.  Next
we will examine the size of the nonlinear effect in the time evolution
by including a mean-field type resummation in the equation.

Our analyses should be useful to grasp deeper understanding of the
qualitative feature inherent to the Glasma~\cite{Lappi:2006fp}, a
transient stage from the coherent CGC state decaying to the
thermalized plasma state.  From our study two remarks are here in
order.
1) Although the initial energy at the collision point in the MV model
contains the UV divergence~\cite{Kovchegov:2005ss,Lappi:2006hq}, the
momentum spectrum of the energy content and the gluon distribution is
well-defined as we will explicitly compute.  The spectrum is quite
informative on its own.
2) Another potential application is the analytical approach toward the
Glasma instability~\cite{Fukushima:2007ja,Fujii:2008dd,Iwazaki:2007es}
found in the numerical simulation~\cite{Romatschke:2005pm}.  Because
the instability presumably resides in the very early time where our
description of the time evolution works, it is a feasible strategy to
examine the stability of field fluctuations on top of the perturbative
time evolution.  We will list more future outlooks in the final
section.

This paper is organized as follows:  In the next section we introduce
the MV model.  Then, in Sec.~\ref{sec:gaussian}, we calculate the
correlation functions of the gluon fields and fix the MV model
parameter corresponding to $Q_s$ for a given infrared regulator.
Section~\ref{sec:steps} is devoted to a guide for our calculation
procedures which are divided into the following four sections;  the
proper time expansion is discussed in Sec.~\ref{sec:proper}, the
energy density and the gluon distribution at $\tau=0$ are evaluated in
Sec.~\ref{sec:initial}, the time evolution is convoluted in
Sec.~\ref{sec:evolution}, and the time evolution is augmented with
nonlinear terms in Sec.~\ref{sec:improve}.  Our discussions and
outlooks are in Sec.~\ref{sec:discussions}.

%%%%%%%%%%   MODEL   %%%%%%%%%%

\section{MODEL}

For the purpose of describing a longitudinally expanding system it is
convenient to adopt the Bjorken coordinates of the proper time $\tau$
and the space-time rapidity $\eta$, defined respectively by
\begin{equation}
 \tau = \sqrt{t^2-z^2}\,,\qquad
 \eta = \frac{1}{2}\ln\biggl[\frac{t+z}{t-z}\biggr]\,.
\end{equation}
It should be noted that $\eta$ above is different from the
pseudo-rapidity in momentum space which is often denoted by $\eta$ in
literature.  The metric tensor associated with the Bjorken coordinates
is $g_{\tau\tau}=1$, $g_{\eta\eta}=-\tau^2$, $g_{xx}=g_{yy}=-1$, and
zero otherwise.

We will work in the radial gauge, $A_\tau=0$, throughout this paper.
The canonical momenta (chromo-electric fields) in this gauge read from
the Lagrangian
as~\cite{Krasnitz:1998ns,Krasnitz:2001qu,Romatschke:2005pm,Fukushima:2006ax}
\begin{align}
 E^i &= \frac{\delta(\tau\Lag)}{\delta(\partial_\tau A_i)}
      = \tau\partial_\tau A_i \,,
\label{eq:mom-i}\\
 E^\eta &= \frac{\delta(\tau\Lag)}{\delta(\partial_\tau A_\eta)}
        = \frac{1}{\tau}\partial_\tau A_\eta \,.
\label{eq:mom-eta}
\end{align}
Here we include $\tau$ in front of $\Lag$ in Eqs.~(\ref{eq:mom-i}) and
(\ref{eq:mom-eta}) originating from $\sqrt{|g|}$ in the integral
measure.  The following equations of motion are derived from
Hamilton's equations:
\begin{align}
 \partial_\tau E^i &= -\frac{\delta(\tau\Ham)}{\delta A_i}
                  = \frac{1}{\tau} D_\eta F_{\eta i}+\tau D_j F_{ji} \,,
\label{eq:eom-i}\\
 \partial_\tau E^\eta &= -\frac{\delta(\tau\Ham)}{\delta A_\eta}
                    = \frac{1}{\tau} D_j F_{j\eta} \,
\label{eq:eom-eta}
\end{align}
with the Hamiltonian,
\begin{equation}
 \Ham = \tr\biggl[\frac{1}{\tau^2}E^i E^i + E^\eta E^\eta
       + \frac{1}{\tau^2}B^i B^i + B^\eta B^\eta \biggr] \,.
\label{eq:hamiltonian}
\end{equation}
These four equations (\ref{eq:mom-i})--(\ref{eq:eom-eta}) are the
basic ingredients for the classical description valid in the early
stage when small-$x$ partons are abundant and quantum corrections are
still negligible;  in momentum rapidity $Y(=\ln(1/x))$ space the
validity region is bounded as $\ln(1/\alpha_s)\ll Y\ll 1/\alpha_s$
specifically.

It has been well established that the initial condition is uniquely
determined by boundary matching at the singularities of the color
source in the limit of vanishing longitudinal extent due to the
Lorentz
contraction~\cite{Kovner:1995ja,Gyulassy:1997vt,Fukushima:2006ax}.
The initial fields are thus known as
\begin{equation}
 \begin{split}
 & A_{i(0)} = \alpha^{(1)}_i + \alpha^{(2)}_i \;,\qquad
   A_{\eta(0)} = 0 \,,\\
 & E^i_{(0)} = 0 \;,\qquad
   E^\eta_{(0)} = ig\bigl[\alpha^{(1)}_i,\alpha^{(2)}_i\bigr] \,,
 \end{split}
\label{eq:initial}
\end{equation}
where $\alpha^{(1)}_i$ and $\alpha^{(2)}_i$ are the \textit{pure}
gauge fields in the space-like region extending from the right-moving
nucleus traveling on the $x^+$ axis and the left-moving nucleus
traveling on the $x^-$ axis, respectively.  They are the gauge
transformation from the light-cone solution~\cite{Kovchegov:1996ty};
\begin{equation}
 \begin{split}
 \alpha^{(1)}_i(\xt) &= -\frac{1}{ig}V_\infty(\xt) \,\partial_i\,
   V^\dagger_\infty(\xt) \,,\\
 \alpha^{(2)}_i(\xt) &= -\frac{1}{ig}W_\infty(\xt) \,\partial_i\,
   W^\dagger_\infty(\xt)
 \end{split}
\label{eq:gauge}
\end{equation}
with the Wilson lines, which are color $\mathrm{SU}(N_c)$ matrices in
the fundamental representation, defined by
\begin{equation}
 \begin{split}
 V^\dagger_{x^-}(\xt) &= \Pcal_- \exp\biggl[-ig\int_{-\infty}^{x^-} \!dz^-\,
  \frac{\rho^{(1)}(\xt,z^-)}{\del^2}\biggr] \,,\\
 W^\dagger_{x^+}(\xt) &= \Pcal_+ \exp\biggl[-ig\int_{-\infty}^{x^+} \!dz^+\,
  \frac{\rho^{(2)}(\xt,z^+)}{\del^2}\biggr] \,.
 \end{split}
\label{eq:vw}
\end{equation}
Here $\Pcal_{\pm}$ represents the path ordering with respect to
$x^\pm$.  We remark that $\rho^{(1)}$ is a static color source in
$x^+$ (i.e.\ $x^+$-independent) and $\rho^{(2)}$ is static in $x^-$
due to the Lorenz time dilatation.  Although we assume
$\rho^{(1)}(\xt,x^-)\propto\delta(x^-)$ and
$\rho^{(2)}(\xt,x^+)\propto\delta(x^+)$ in the end, we have to keep
both the $z$-integral and the path ordering, which encompasses the
random color distribution along the longitudinal
extent~\cite{Fukushima:2007ki,review,Gelis:2001da,Blaizot:2004wu,Blaizot:2004wv,Fukushima:2007dy}.

Now that we have fixed the initial condition for the equations of
motion, we should have a unique solution.  In principle the solution
is given in terms of $V$ and $W$ through the initial condition, and
any physical observable $\mathcal{O}$ at later time could be
determined once $\rho^{(1)}$ and $\rho^{(2)}$ are given.  Since the
color sources $\rho^{(1)}$ and $\rho^{(2)}$ will fluctuate randomly in
event by event, observed $\mathcal{O}$ should be an ensemble average
over the color source distribution
$\mathcal{W}[\rho^{(1)}\!,\rho^{(2)}]$, that is,
\begin{equation}
 \bigl\langle\mathcal{O}\bigr\rangle = \int\D\rho^{(1)}\D\rho^{(2)}\,
  \mathcal{W}[\rho^{(1)}\!,\rho^{(2)}]\,\mathcal{O}[V,W] \,.
\end{equation}
This is a general form for the observable expectation value in the CGC
framework.

The MV model is founded on the Gaussian approximation to
$\mathcal{W}[\rho^{(1)}\!,\rho^{(2)}]$ as formulated in
Ref.~\cite{Iancu:2002aq}, where any averaged quantity is expressed in
terms of the two-point function,
\begin{equation}
 \begin{split}
 &\bigl\langle\rho^{(m)a}(\xt,z)\rho^{(n)b}(\yt,z')\bigr\rangle \\
 & = g^2\mu^2(z)\,\delta^{mn}\,\delta^{ab}\,\delta(z-z')\,
  \delta^{(2)}(\xt\!-\yt) \,.
 \end{split}
\label{eq:gauss}
\end{equation}
Here $m$ and $n$ distinguish the right-moving (1) and left-moving (2)
nucleus, $a$ and $b$ are the color indices, and $\mu^2(z)$ is the only
dimensional parameter in the MV model.  All the physical quantities
are given in terms of (integrated) $\mu^2$ and importantly $\mu^2$ has
a tight connection to the parton saturation scale $Q_s^2$.

With these preliminaries, we are ready to compute physical quantities
of our interest in the MV model, once the model scale $\mu^2$ and the
coupling constant $g$ are fixed.  Before proceeding to the next
section, let us elucidate a simple expression for $\alpha^{(n)}$ for
later use.  Here we only deal with $\alpha^{(1)}$ of the right-moving
nucleus and suppress the superscript in the rest of this section and
in the next section.  Similar formulae obviously hold for
$\alpha^{(2)}$, too.

Substituting Eq.~(\ref{eq:vw}) into Eq.~(\ref{eq:gauge}), we can
explicitly write down the expression,
\begin{equation}
 \begin{split}
  \alpha_i &= -\frac{1}{ig}\int_{-\infty}^{\infty}\! dx^-
   V_{x^-} \biggl[-ig\frac{\partial_i\rho(x^-)}{\del^2}\biggr]
   V^\dagger_{x^-} \\
  &= \int_{-\infty}^{\infty}\! dx^-
   \frac{\partial_i\rho^a(x^-)}{\del^2}\,
   V_{x^-}\, \tf^a\, V^\dagger_{x^-} \\
  &= \int_{-\infty}^{\infty}\! dx^-
   \frac{\partial_i\rho^a(x^-)}{\del^2}\,
   \Vad^{\dagger ab}_{x^-}\, \tf^b \,,
 \end{split}
\label{eq:s-gauge}
\end{equation}
where $\tf^a$'s are the $\mathrm{SU}(N_c)$ algebra in the fundamental
representation and $\Vad$ denotes the Wilson line in the adjoint
representation whose components are given by
$\Vad^{ab}=2\tr[\tf^a V \tf^b V^\dagger]$ in terms of the fundamental
Wilson lines.

%%%%%%%%%%   GAUSSIAN AVERAGE   %%%%%%%%%%

\section{GAUSSIAN AVERAGE}
\label{sec:gaussian}

In this section we list the useful formulae for the Gaussian average
over the color source distribution.  Using Eq.~(\ref{eq:s-gauge}) we
can explicitly write the gauge field correlator in a form of
\begin{align}
 & \Bigl\langle \alpha_i^a(\xt)\,\alpha_j^b(\yt)
  \Bigr\rangle \notag\\
 &= \int_{-\infty}^{\infty}\! dx^- dy^- \biggl\langle
  \frac{\partial_i\rho^c(\xt,x^-)}{\del^2_x}\!\cdot\!
  \frac{\partial_j\rho^{c'}(\yt,y^-)}{\del^2_y} \Bigr\rangle \notag\\
 &\quad \times \Bigl\langle\Vad^{\dagger ca}_{x^-}(\xt)
   \Vad^{\dagger c'b}_{y^-}(\yt) \Bigr\rangle \notag\\
 &= \delta^{ab} g^2\int_{-\infty}^\infty\! dx^-\, \partial_i^x
  \partial_j^y L(\xt\!-\yt) \,\mu^2(x^-) \notag\\
 &\quad\times C_{\text{adj}}(x^-;\xt\!-\yt)\,,
\label{eq:alpha-cor}
\end{align}
where we have defined
\begin{equation}
 \begin{split}
 & \biggl\langle \frac{\rho^a(\xt,x^-)}{\del^2_x}\!\cdot\!
  \frac{\rho^b(\yt,y^-)}{\del^2_y} \biggr\rangle \\
 &= \delta^{ab} g^2\mu^2(x^-)\,\delta(x^-\!-y^-)
  \,L(\xt\!-\yt) \,,
 \end{split}
\label{eq:rho-cor}
\end{equation}
and
\begin{equation}
 L(\xt) = \frac{1}{(\del^2)^2}\,\delta^{(2)}(\xt) \\
 = \int \frac{d^2\kt}{(2\pi)^2}\,\frac{e^{i\kt\cdot\xt}}
  {(\kt^2+m^2)^2} \,.
\label{eq:gamma}
\end{equation}
The momentum integration in Eq.~(\ref{eq:gamma}) has infrared (IR)
divergence.  We regularize this by the gluon mass $m\neq0$ and regard
$m$ as a model parameter controlling the IR behavior.  Then the
integration can be performed analytically.  For convenience we
introduce the notation~\cite{Gelis:2001da,Blaizot:2004wu},
$\Gamma(\xt)=2L(\zero)-2L(\xt)$, which appears in the expressions for
the physical quantities and is expressed as
\begin{equation}
 \begin{split}
 \Gamma(\xt) &= \frac{1}{2\pi m^2}\Bigl[1- m|\xt|\,K_1(m|\xt|)\Bigr] \\
 &\simeq -\frac{|\xt|^2}{8\pi}\ln\bigl[ m^2|\xt|^2 \lambda^2\bigr] \,,
 \end{split}
\label{eq:approx}
\end{equation}
where $K_1(x)$ is the modified Bessel function, and
$\lambda=e^{\gamma_E-1/2}/2\simeq 0.54$.  The second line is an
approximate expression for small $m|\xt|$.

Besides IR singularity, UV divergence also exists in $\del^2L(\xt)$,
$(\del^2)^2 L(\xt)$ and so on, due to the Delta function correlation
$\delta^{(2)}(\xt)$ in Eq.~(\ref{eq:gauss}), as explicitly shown below
in this section.

In Eq.~(\ref{eq:alpha-cor}), we have also defined the Wilson line
correlator by
\begin{equation}
 \Bigl\langle\Vad^{\dagger ca}_{x^-}(\xt)
  \Vad^{\dagger cb}_{x^-}(\yt) \Bigr\rangle 
 = \delta^{ab} C_{\text{adj}}(x^-;\xt\!-\yt) \,.
\label{eq:V-cor}
\end{equation}
We note that Eq.~(\ref{eq:rho-cor}) is the building block and its
iterative use in the left-hand side of Eq.~(\ref{eq:V-cor}) leads to
$C_{\text{adj}}(x^-;\xt)$.  The answer is already given in
literature~\cite{Gelis:2001da,Blaizot:2004wu,Fukushima:2007dy} as
\begin{equation}
 C_{\text{adj}}(x^-;\xt)
 = \exp\biggl[ -\frac{N_c}{2}\,
  g^4 \int^{x^-}_{-\infty}\!dz\,
  \mu^2(z)\,\Gamma(\xt) \biggr] \,.
\end{equation}
This factor encodes the effect of the multiple interactions with the
color source in the nucleus.  We see that
$C_{\text{adj}}(x^-;\xt\!\to\!\zero)\to 1$ as it should be due to
color transparency.

To simplify the discussion we assume here
$\mu^2(x^-)=\theta(x^-)\theta(\epsilon-x^-)\mu_A^2/\epsilon$ with
$\epsilon\to0^+$ (this can be relaxed in more careful calculations) so
that
\begin{equation}
 \int_{-\infty}^\infty \!dx^-\, \mu^2(x^-) = \mu^2_A \,.
\end{equation}
Then the integration over $x^-$ results in
\begin{equation}
 \int_{-\infty}^{\infty} \!dx^- \mu^2(x^-)\,C_{\text{adj}}(x^-;\xt)
 = \mu_A^2\,\bar{C}_{\text{adj}}(\xt) 
\end{equation}
with
\begin{equation}
 \bar{C}_{\text{adj}}(\xt) = \frac{1}{a m^2\Gamma(\xt)} \biggl(
  1-\exp\Bigl[-a m^2\Gamma(\xt) \Bigr]\biggr) \,.
\end{equation}
We have used a dimensionless parameter:
\begin{equation}
 a \equiv \frac{\Nc(g^2\mu_A)^2}{2m^2} \,,
\label{eq:alpha}
\end{equation}
in accord with
Refs.~\cite{Gelis:2001da,Blaizot:2004wu,Blaizot:2004wv}.  Here in this
paper, however, we regard the parameter $a$ as the IR parameter
controlling how much of the non-perturbative effect in the initial
fields contributes to the momentum integration once the saturation
scale is fixed [see discussions below].  Finally we find from
Eq.~(\ref{eq:alpha-cor}),
\begin{equation}
 \begin{split}
 &\Bigl\langle \alpha^a_i(\xt)\,\alpha^b_j(\yt)
  \Bigr\rangle \\
 &= \delta^{ab} g^2\mu_A^2\, \bar C_{\text{adj}}(\xt\!\!-\!\yt)\,
  \partial_i^x\partial_j^y L(\xt\!\!-\yt) \,.
 \end{split}
\end{equation}

Let us discuss a rough estimate for the parameter $a$ here.  We will
check at the end of this section that $g^2\mu_A$ is nearly identified
with the saturation scale $Q_s$ (which also depends on $a$).  The IR
cutoff is difficult to fix uniquely but a natural expectation is
$m\simeq g\mu_A$, from the comparison between our cutoff scheme and the
color neutrality as a result of the quantum
evolution~\cite{Iancu:2002aq,Mueller:2002pi} with an uncertain
prefactor.  The strong coupling constant at the scale $Q_s$ at RHIC
energy is $\alpha_s\simeq0.3$ (or $g\simeq 2$).  This implies
$a\simeq 3\cdot g^2 /2=6$.  The physical meaning of this $m$ value is
that the classical description breaks down there.  On the other hand,
if we take $m$ of order of the confinement scale $\sim 1\fm$, i.e.,
$m\simeq 200\MeV$, we would have
$a \simeq (3\cdot 1\sim 2\GeV^2)/(2\cdot 0.2^2\GeV^2)=37\sim 75$ for
$Q_s^2=1\sim 2\GeV^2$.  In this estimate $m$ is the scale where the
CGC picture in the perturbative regime breaks down.  From these
considerations the reasonable range for $a$ would be $10$--$100$ at
the RHIC energy.  It should be noted that $a$ used in the numerical
simulation may be much larger;  there $m$ is provided by the system
size of order $10\fm$ and $a$ may be then hundreds times larger.
Nevertheless, because the IR property in the numerical implementation
of the MV model is totally different from the analytical
formulation~\cite{Fukushima:2007ki}, we cannot make a direct
comparison to the numerical results.  In this paper we will
specifically choose four cases:  $a=10$, $25$, $100$, and $500$ for
comparison, keeping in mind that the physical value is around
$a=10\sim 40$.

It is important to note that the above $\bar{C}_{\text{adj}}(\xt)$
exactly corresponds to the gluon propagator~\cite{Kovchegov:1996ty}
and should be distinguished from
\begin{equation}
 C_A(\xt)=\exp\bigl[-a m^2\Gamma(\xt)\bigr] \,,
\end{equation}
which appears in the gluon
distribution~\cite{Gelis:2001da,Blaizot:2004wu,Lappi:2007ku}.  In the
case of dense-dilute collisions it is possible to convert
$\bar{C}_{\text{adj}}(\xt)$ into $C_A(\xt)$ by gauge rotation but this
is not the case if the dense-dense collisions are concerned in a
symmetric way.

\begin{figure}
 \includegraphics[width=8cm]{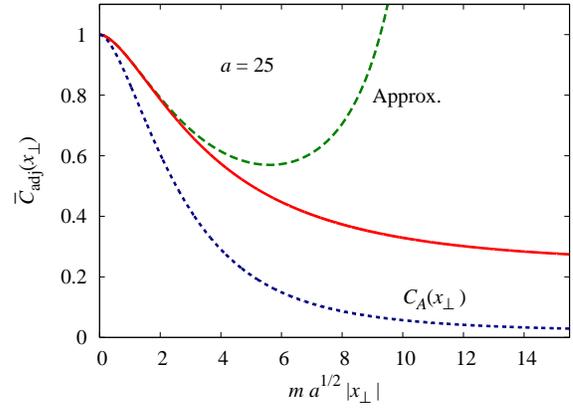}
 \caption{$\bar{C}_{\text{adj}}(\xt)$ as a function of dimensionless
   $m\sqrt{a}|\xt|=\sqrt{\Nc/2}\,g^2\mu_A |\xt|$ in the case that
   $a=\Nc(g^2\mu_A)^2/(2m^2)$ is $25$ which is the same choice as
   Ref.~\cite{Blaizot:2004wu}.  The solid curve is the result from the
   numerical integration and the dashed curve represents the
   approximation by Eq.~(\ref{eq:approx}) which eventually blows up in
   the IR region.}
 \label{fig:cx}
\end{figure}

The part involving $\bar{C}_{\text{adj}}(\xt)$ describes the
transverse color correlation in the random distribution.  It shows
moderate damping in a distant region as plotted by the (red) solid
curve in Fig.~\ref{fig:cx} for $a=25$.  We show the approximation
using Eq.~(\ref{eq:approx}) by the (green) dashed curve, and
$C_A(\xt)$ by the (blue) dotted curve for reference.  From the
comparison between the solid and dotted curves we immediately realize
that $\bar{C}_{\text{adj}}(\xt)$ retains a long-range tail than
$C_A(\xt)$.  Indeed, we can readily notice that
$\Gamma(\xt)\to (2\pi m^2)^{-1}$ for $m|\xt|\gg 1$ and thus
\begin{equation}
 \bar{C}_{\text{adj}}(\xt) \to \frac{2\pi}{a}
  \bigl( 1-e^{-a/2\pi} \bigr) \,,
\end{equation}
which is nearly $0.50$, $0.25$, $0.063$, and $0.013$ for $a=10$,
$25$, $100$, and $500$, respectively;  the asymptotic value is fixed
by the IR cutoff $a$.

Equation~(\ref{eq:alpha-cor}) is a useful expression.  Here we
enumerate quantities necessary for the evaluation of the initial
energy;  the equal-point gauge field correlator is
\begin{equation}
 \Bigl\langle\alpha^a_i(\xt)\alpha^b_j(\xt)\Bigr\rangle
 = -\frac{1}{2}\,\delta^{ab}\delta_{ij}\,
  g^2\mu_A^2\, \del^2L(\zero) \,,
\label{eq:corr}
\end{equation}
and the derivative correlations read
\begin{align}
 &\Bigl\langle(\partial_k\alpha^a_i(\xt))(\partial_l\alpha^b_j(\xt))
  \Bigr\rangle \notag\\
 &= \frac{1}{8}\,\delta^{ab}\,g^2\mu_A^2\biggl[(\delta_{ij}\delta_{kl}
  +\delta_{ik}\delta_{jl}+\delta_{il}\delta_{jk})\, (\del^2)^2L(\zero) \notag\\
 &\qquad -N_c g^4\mu_A^2\,\delta_{ij}\delta_{kl}
  (\del^2L(\zero))^2 \biggr] \,,
\end{align}
and also
\begin{align}
 &\Bigl\langle \alpha^a_i(\xt)\del^2\alpha^b_j(\xt)\Bigr\rangle 
 = -\frac{1}{4}\,\delta^{ab}\,g^2\mu_A^2\biggl[2\delta_{ij}
  (\del^2)^2L(\zero) \notag\\
 &\qquad -N_c g^4\mu_A^2\,\delta_{ij}
  (\del^2L(\zero))^2 \biggr] \,,
\end{align}
where we can explicitly estimate the integrals as
\begin{equation}
 \del^2L(\zero) = -\int^\Lambda \frac{d^2\kt}{(2\pi)^2}
  \frac{1}{\kt^2+m^2} = -\frac{1}{2\pi}\ln\frac{\Lambda}{m} \,,
\label{eq:log}
\end{equation}
and
\begin{equation}
 (\del^2)^2L(\zero) = \int^\Lambda \frac{d^2\kt}{(2\pi)^2}
  = \frac{1}{4\pi}\Lambda^2
\end{equation}
for $\Lambda\gg m$.  We see that the derivatives of the correlation
$L(\xt)$ at the origin contain the UV divergence.

Finally in this section, let us make the connection between the scale
parameter, $g^2 \mu_A$, and the saturation scale $Q_s$.  The
definition of the saturation scale is not unique.  We apply the same
scheme as adopted in Ref.~\cite{Lappi:2007ku} using the Fourier
transform of $C_A(\xt)$, which we denote as $C_A(\kt)$.  Since
$k_\perp^2 C_A(\kt)$ is interpreted as a gluon distribution function,
we define the saturation scale as the peak position of this function.
Bear in mind that we have to keep the same definition for $Q_s$ in
order to make a consistent comparison with other empirical analyses.
[$Q_s$ itself is not gauge invariant.]

\begin{figure}
 \includegraphics[width=8cm]{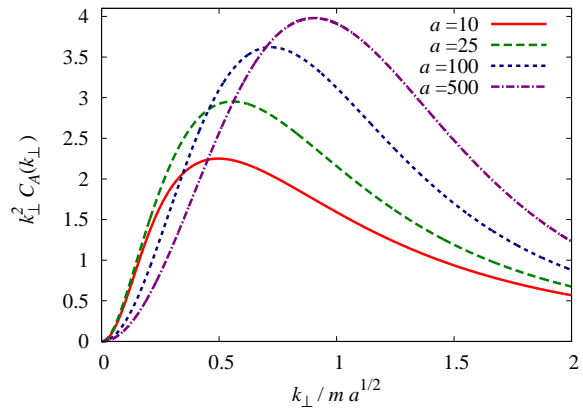}
 \caption{$k_\perp^2 C_A(\kt)$ as a function of the dimensionless
   variable $k_\perp/(m\sqrt{a})$ where
   $m\sqrt{a}=\sqrt{\Nc/2}\,g^2\mu_A$ for $a=10$, $25$,
   $100$, and $500$.}
 \label{fig:ck}
\end{figure}

Figure~\ref{fig:ck} is a plot for $k_\perp^2 C_A(\kt)$ as a function
of $k_\perp/(m\sqrt{a})$ where $m\sqrt{a}=\sqrt{\Nc/2}\,g^2\mu_A$.
The peak position defines the saturation scale.  From
Fig.~\ref{fig:ck} it reads $Q_s/(m\sqrt{a})=0.497$, $0.558$, $0.720$,
and $0.903$ for $a=10$, $25$, $100$, and $500$, which correspond to
$m=0.64Q_s$, $0.36Q_s$, $0.14Q_s$, and $0.050Q_s$, respectively.  We
summarize these relations in Table~\ref{tab:saturation}.

\begin{table}
 \begin{tabular}{|c||c|c|}
\hline
 $a$ & $m$ & $g^2\mu_A$ \\
\hline
 $\quad 10\quad$ & $\quad 0.64Q_s\quad$ & $\quad 1.65Q_s\quad$ \\
\hline
 $25$ & $0.36Q_s$ & $1.46Q_s$ \\
\hline
 $100$ & $0.14Q_s$ & $1.13Q_s$ \\
\hline
 $500$ & $0.050Q_s$ & $0.90Q_s$ \\
\hline
 \end{tabular}
 \caption{IR cutoff $m$ and the MV model parameter $g^2\mu_A$
   determined from $k_\perp^2 C_A(\kt)$ for various $a$.}
 \label{tab:saturation}
\end{table}

%%%%%%%%%%   CALCULATION STEPS   %%%%%%%%%%

\section{CALCULATION STEPS}
\label{sec:steps}

In the subsequent sections we will elucidate the time evolution of the
initial energy density and the gluon distribution.  Since the
computation process is involved, we outline the derivation in advance
here.

{\bf Step I)}  We will carry out the proper time expansion;  the
equations of motion are solved in a power series of $\tau$.

{\bf Step II)}  We will take the Gaussian average over the color
source distribution order by order of $\tau$.  We will then find that
each term contains the UV singularity.  We decompose the divergent
initial energy density into the Fourier modes which give the gluon
distribution if divided by the gluon energy.  We take full account of
the saturation effect.

{\bf Step III)}  We will figure out the time evolution for the UV
modes from the equations of motion.  We confirm that the UV
singularity is regularized at finite $\tau$, and fix the initial
condition for the time evolution by matching it to Step II).

{\bf Step IV)}  We will improve the time evolution by inclusion of the
nonlinear terms in the Gaussian approximation.  We finally read the
initial energy density and the gluon number at the formation time when
the time dependence shows free-streaming behavior.

%%%%%%%%%%   PROPER TIME EXPANSION  -- Step I)   %%%%%%%%%%

\section{PROPER TIME EXPANSION -- Step I)}
\label{sec:proper}

We shall begin with a naive expansion for the classical fields in
terms of the proper time $\tau$ as in Ref.~\cite{Fries:2006pv}, and
in the next section we resum the leading order terms in the UV
singularity coming from the Gaussian average.  It should be stressed
that the gauge fields for a given color source are free from the
singularity and have no difficulty in the $\tau$-expansion.

The first order terms are the contributions at $\tau=0$, that is, the
initial fields as given in Eq.~(\ref{eq:initial}).  Let us now
evaluate the associated chromo-electric and chromo-magnetic fields.
It is trivial to read the initial chromo-electric fields from the
initial condition as
\begin{equation}
 \begin{split}
 E^i_{(0)} &= 0 \,,\\
 E^\eta_{(0)} &= ig\Bigl( \bigl[\alpha^{(1)}_1,\alpha^{(2)}_1\bigr]
  +\bigl[\alpha^{(1)}_2,\alpha^{(2)}_2\bigr] \Bigr) \,,
 \end{split}
\label{eq:initial-electric}
\end{equation}
and noting $\alpha^{(n)}_i$'s are pure gauge solutions, we find the
chromo-magnetic fields being 
\begin{equation}
 B^i_{(0)} = 0 \,,\qquad
 B^\eta_{(0)} = F_{12(0)} \,,
\label{eq:initial-magnetic}
\end{equation}
where
\begin{equation}
 F_{ij(0)} = -ig\Bigl( \bigl[\alpha^{(1)}_i,\alpha^{(2)}_j\bigr]
  +\bigl[\alpha^{(2)}_i,\alpha^{(1)}_j\bigr] \Bigr) \,.
\end{equation}
These results are recognized as the fundamental property of the Glasma
initial state;  the longitudinal fields between two color sheets are
predominant in the initial state of matter.

To proceed further away from $\tau=0$, we apply the expansion, i.e.,
\begin{equation}
 \mathcal{O}(\tau) = \sum_{n=0}^\infty \mathcal{O}_{(n)}\tau^n \,,
\end{equation}
for arbitrary fields $\mathcal{O}$ given in terms of the classical
gauge fields.  It is easy to confirm that the terms for odd $n$ are
all vanishing due to time reversal symmetry.  The non-trivial
contribution to the gauge field starts at the second-order terms which
turn out to be
\begin{equation}
 \begin{split}
 & A_{i(2)} = \half E^i_{(2)} = \quart D_{j(0)}F_{ji(0)} \,,\\
 & A_{\eta(2)} = \half E^\eta_{(0)} \,.
 \end{split}
\end{equation}
Then we can explicitly write down the second-order contributions to
the initial chromo-electric and chromo-magnetic fields:
\begin{equation}
 \begin{split}
 E^i_{(2)} &= \half D_{j(0)}F_{ji(0)}
  = -\epsilon^{ij}\half D_{j(0)}B^\eta_{(0)}\,,\\
 E^\eta_{(2)} &= \half D_{j(0)}F_{j\eta(2)} = \quart D_{j(0)}D_{j(0)}E^\eta_{(0)} \,,
 \end{split}
\label{eq:e_second}
\end{equation}
and
\begin{equation}
 \begin{split}
 B^i_{(2)} &= \epsilon^{ij}F_{j\eta(2)}
  = \epsilon^{ij} \half D_{j(0)}E^\eta_{(0)} \,,\\
 B^\eta_{(2)} &= F_{12(2)} = \quart D_{j(0)}D_{j(0)}B^\eta_{(0)} \,.
 \end{split}
\label{eq:b_second}
\end{equation}
Here we have defined the anti-symmetric tensor
$\epsilon^{12}=-\epsilon^{21}=1$ in the transverse coordinates.  The
duality relation between the electric and magnetic fields is manifest
itself in Eqs.~(\ref{eq:e_second}) and (\ref{eq:b_second}).

%%%%%%%%%%   INITIAL ENERGY DENSITY AND GLUON DISTRIBUTION -- Step II)  %%%%%%%%%%

\section{INITIAL ENERGY DENSITY AND GLUON DISTRIBUTION -- Step II)}
\label{sec:initial}

The energy density is an ensemble average of the Hamiltonian
(\ref{eq:hamiltonian}), i.e.\ $\energy=\langle\Ham\rangle$.   In this
section we compute $\energy$ and decompose it into the Fourier
components.  The merit of the Fourier decomposition is that each
Fourier mode is free from UV singularity and we realize how the
$\kt$-integration of the spectrum diverges.  Besides, we can define
the gluon distribution associated with the field intensity at each
momentum.

%%%   Zeroth Order

\subsection{Zeroth Order}

Right after the collision at $\tau=0$ only the longitudinal fields
along the $\eta$ direction have nonzero values, and therefore we
calculate the following energy density:
\begin{equation}
 \energy_{(0)} = \Bigl\langle \tr\bigl[ E^\eta_{(0)} E^\eta_{(0)}
  + B^\eta_{(0)} B^\eta_{(0)} \bigr] \Bigr\rangle \,.
\label{eq:e0}
\end{equation}
Noting that $\Gamma(\xt\!-\yt)\to 0$ in the limit of $\yt\to\xt$,
we can drop $\bar{C}_{\text{adj}}(\xt\!-\!\yt)$ in Eq.~(\ref{eq:e0}).
With a shorthand notation as
$\langle\alpha_i^{(m)a}\alpha_j^{(n)b}\rangle =\delta^{mn}\delta^{ab}
\delta_{ij}\langle\alpha\alpha\rangle$, we can express the energy
density in a form of
\begin{equation}
 \energy_{(0)} = 2\Nc(\Nc^2-1)g^2\langle\alpha\alpha\rangle^2
  = g^6\mu^4_A \cdot\frac{3}{\pi^2}\biggl[\ln\frac{\Lambda}{m}
  \biggr]^2 \,,
\label{eq:energy}
\end{equation}
from Eq.~(\ref{eq:log}) for $\Nc=3$.

It is interesting and practically useful to look into the energy
content in momentum space, which is also necessary when we consider
the gluon distribution.  To this end, let us calculate the following
quantity:
\begin{equation}
 \begin{split}
 \energy_{(0)}(\kt) &= \frac{1}{V}\Bigl\langle \tr\bigl[
  E^\eta_{(0)}(-\kt) E^\eta_{(0)}(\kt) \\
 &\qquad\qquad + B^\eta_{(0)}(-\kt) B^\eta_{(0)}(\kt) \bigr]
  \Bigr\rangle \,,
 \end{split}
\label{eq:tilenergy}
\end{equation}
so that the $\kt$-integration of $\energy_{(0)}(\kt)$ recovers the
energy density $\energy_{(0)}$.  Then we have to evaluate the
spatially separated correlation function of the
chromo-electric and chromo-magnetic fields:
\begin{equation}
 \begin{split}
 &\Bigl\langle\tr E^\eta_{(0)}(\xt)E^\eta_{(0)}(\yt) \Bigr\rangle
  = \frac{1}{2}\Nc(\Nc^2-1)g^2 \\
 &\times \Bigl( \langle\alpha_1(\xt)\alpha_1(\yt)\rangle^2
  +\langle\alpha_1(\xt)\alpha_2(\yt)\rangle^2 \\
 &\qquad +\langle\alpha_2(\xt)\alpha_1(\yt)\rangle^2
  +\langle\alpha_2(\xt)\alpha_2(\yt)\rangle^2 \Bigr) \\
 &= \frac{1}{2}\Nc(\Nc^2-1)g^6\mu_A^4
  \bigl[\bar{C}_{\text{adj}}(\xt\!-\yt)\bigr]^2\\
 &\quad\times \Bigl[ (\partial_1^2L(\xt\!-\yt))^2
  +2(\partial_1\partial_2L(\xt\!-\yt))^2 \\
 &\qquad\qquad + (\partial_2^2L(\xt\!-\yt))^2 \Bigr] \,.
 \end{split}
\label{eq:EE0}
\end{equation}
Computing the chromo-magnetic part in the same way, we obtain almost
the same expression as Eq.~(\ref{eq:EE0}) but with the opposite sign in
front of the $[\partial_1\partial_2L(\xt\!-\!\yt)]^2$ term.  This term
cancels out in the sum of the chromo-electric and chromo-magnetic
parts.  The Fourier transformation of the sum gives rise to
\begin{equation}
 \begin{split}
 &\energy_{(0)}(k_\perp)
  = \frac{1}{V}\int d^2\xt d^2\yt \,e^{-i\kt(\xt-\yt)} \\
 &\qquad \times \Bigl\langle \tr\bigl[ E^\eta_{(0)}(\xt)E^\eta_{(0)}(\yt)
  +B^\eta_{(0)}(\xt)B^\eta_{(0)}(\yt) \bigr] \Bigr\rangle \\
 &= \frac{1}{2}\Nc(\Nc^2-1)g^6\mu_A^4 \\
 &\times \int\frac{d^2\qt{1}}{(2\pi)^2}\frac{d^2\qt{2}}{(2\pi)^2}
  \frac{d^2\qt{3}}{(2\pi)^2} \frac{\bar{C}_{\text{adj}}(\qt{1})
  \bar{C}_{\text{adj}}(\qt{2})}
  {(\qt{3}^2\!\!+m^2)[(\kt'\!\!-\!\qt{3})^2+m^2]} \,.
 \end{split}
\end{equation}
We denote $\kt'=\kt\!-\qt{1}\!-\qt{2}$, and
$\bar{C}_{\text{adj}}(\qt{})$ is the Fourier transform of
$\bar{C}_{\text{adj}}(\xt)$.  We can further perform the
$\qt{3}$-integration to reach the following:
\begin{align}
 &\energy_{(0)}(k_\perp) = \frac{1}{4\pi}\Nc(\Nc^2-1)g^6\mu_A^4
  \int\frac{d^2\qt{1}}{(2\pi)^2} \frac{d^2\qt{2}}{(2\pi)^2} \notag\\
 &\quad\times \frac{\bar{C}_{\text{adj}}(\qt{1})\bar{C}_{\text{adj}}(\qt{2})}
  {k_\perp'\sqrt{k_\perp^{\prime2}+4m^2}} \ln\Biggl[
  \frac{\sqrt{k_\perp^{\prime2}+4m^2}+k_\perp'}{\sqrt{k_\perp^{\prime2}
  +4m^2}-k_\perp'}\Biggr] \,.
\label{eq:energy-zero}
\end{align}
At this point it is easy to confirm that Eq.~(\ref{eq:energy-zero})
reproduces Eq.~(\ref{eq:energy}) when integrated over $\kt$.  In the
presence of the $\kt$-integration, we can change the variable from
$\kt$ to $\kt'\!=\!\kt\!-\!\qt{1}\!-\!\qt{2}$ and then we can perform
the $\qt{1,2}$-integrations independently of the $\kt$-integration.
Using the sum rule,
\begin{equation}
 \int\frac{d^2\qt{1,2}}{(2\pi)^2}\, \bar{C}_{\text{adj}}(\qt{1,2})
  = \bar{C}_{\text{adj}}(\xt=\zero) = 1 \,,
\label{eq:sum}
\end{equation}
we have
\begin{align}
 \energy_{(0)} &= \frac{1}{4\pi}\Nc(\Nc^2-1)g^6 \mu_A^4
  \int^\Lambda\frac{d^2 k_\perp}{(2\pi)^2}\, \frac{1}{m^2}\,
  \mathcal{T}_{\text{pert}}(k_\perp/m) \notag\\
 &= \frac{1}{8\pi^2}\Nc(\Nc^2-1)g^6\mu_A^4 \biggl[\ln
  \frac{\Lambda}{m}\biggr]^2 \,,
\label{eq:energy-confirm}
\end{align}
where we have used $\Lambda\gg m$ and this is exactly identical to
Eq.~(\ref{eq:energy}).  Here, we have defined the perturbative
contribution by a function,
\begin{equation}
 \mathcal{T}_{\text{pert}}(\xi) = \frac{1}{\xi\sqrt{\xi^2+4}}\ln
  \biggl[\frac{\sqrt{\xi^2+4}+\xi}{\sqrt{\xi^2+4}-\xi}\biggr] \,.
\end{equation}
It is worth mentioning that
$\mathcal{T}_{\text{pert}}(\xi\to0)\to 0.5$ which is finite 
owing to IR regularization.

\begin{figure}
 \includegraphics[width=8cm]{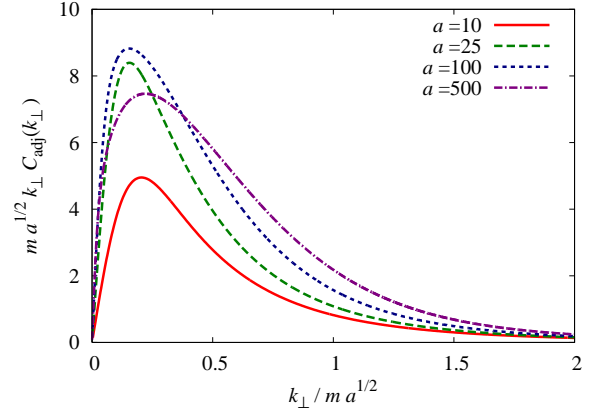}
 \caption{$m\sqrt{a}\,k_\perp\bar{C}_{\text{adj}}(\kt)$ as a
   function of the dimensionless variable $k_\perp/(m\sqrt{a})$
   for $a=10$, $25$, $100$, and $500$.}
 \label{fig:ck2}
\end{figure}

We need to evaluate $\bar{C}_{\text{adj}}(\kt)$ appearing in the
integral (\ref{eq:energy-zero}).  In Fig.~\ref{fig:ck2} we plot
$m\sqrt{a}\,k_\perp \bar{C}_{\text{adj}}(\kt)$ as a function of
$k_\perp/(m\sqrt{a})$.  As is already mentioned previously,
$\bar{C}_{\text{adj}}(\xt)$ has a constant tail 
$(2\pi/a)(1-\exp[-a/(2\pi)])$ at large distances due to the IR cutoff,
which gives rise to a Dirac delta function $\delta^{(2)}(\kt)$ in
momentum space.  $\bar{C}_{\text{adj}}(\kt)$ represents the effect of
the multiple scattering. For smaller $a$ with fixed $g^2 \mu_A$,
larger part of IR region is cutoff and therefore
$\bar{C}_{\text{adj}}(\kt)$ spreads over smaller $\kt$ region and has
larger weight of $\delta^{(2)}(\kt)$.  For larger $a$, more multiple
interactions are effective and $\bar{C}_{\text{adj}}(\kt)$ has the
larger spread and smaller weight of $\delta^{(2)}(\kt)$.  The
perturbative limit corresponds to no multiple scattering.

\begin{figure}
 \includegraphics[width=8cm]{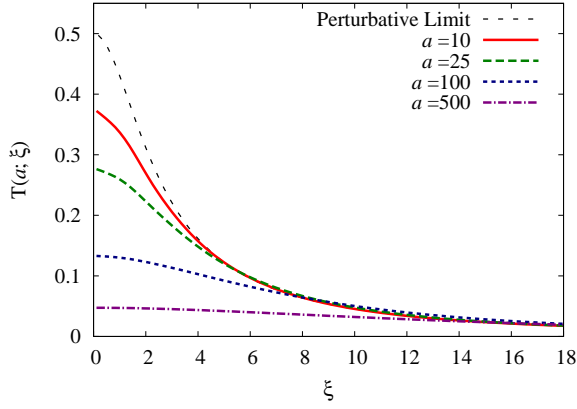}
 \caption{$\mathcal{T}(a ; \xi)$ in the initial energy density as a
   function of the dimensionless variable $\xi(=k_\perp/m)$;  the
   (black) dotted curve represents the perturbative limit
   $\mathcal{T}_{\text{pert}}(\xi)$ (where
   $\bar{C}_{\text{adj}}(\qt{1,2})$ is replaced with
   $(2\pi)^2\delta^{(2)}(\qt{1,2})$), the (red) solid curve for
   $a=10$, the (green) dashed curve for $a=25$, the (blue) dotted
   curve for $a=100$, and the (purple) dot-dashed curve for $a=500$.}
 \label{fig:integ}
\end{figure}

Using the numerical results shown in Fig.~\ref{fig:ck2} and the Dirac
delta function with the weight $(2\pi/a)(1-\exp[-a/(2\pi)])$, we can
explicitly evaluate $\energy_{(0)}(k_\perp)$ performing the numerical
integration of $\qt{1}$ and $\qt{2}$.  In the perturbative limit where
$\bar{C}_{\text{adj}}(\qt{1,2})$ factor is replaced with
$(2\pi)^2\delta^{(2)}(\qt{1,2})$ in Eq.~(\ref{eq:energy-zero}), the
$\kt$-spectrum of the energy content is simply fixed by
$\mathcal{T}_{\text{pert}}(k_\perp/m)$.  We show this perturbative
limit of $\mathcal{T}_{\text{pert}}(k_\perp/m)$ by the (black) dotted
curve in Fig.~\ref{fig:integ}.

The saturation effect changes the functional form of
$\mathcal{T}_{\text{pert}}(\xi)$ into the one depending on $a$ which
we denote by $\mathcal{T}(a ; \xi)$, i.e.,
\begin{equation}
 \energy_{(0)}(k_\perp) = \frac{1}{4\pi m^2}\Nc(\Nc^2-1)g^6 \mu_A^4\,
  \mathcal{T}(a ; k_\perp/m) \,.
\end{equation}
Figure~\ref{fig:integ} shows $\mathcal{T}(10;\xi)$ by the (red) solid
curve, $\mathcal{T}(25;\xi)$ by the (green) dashed curve,
$\mathcal{T}(100;\xi)$ by the (blue) dotted curve, and
$\mathcal{T}(500;\xi)$ by the (purple) dot-dashed curve.  We notice at
a glance at Fig.~\ref{fig:integ} that the saturation effect is larger
for larger $a$ in the low $k_\perp$ region.  Although the functional
form of $\bar{C}_{\text{adj}}(\kt)$ has a large suppression for small
$a$ in view of Fig.~\ref{fig:ck2}, the resulting spectrum in
Fig.~\ref{fig:integ} turns out to be closer to the perturbative one
when $a$ is small.  This is because $\bar{C}_{\text{adj}}(\qt{1,2})$
contains the Dirac delta function $\delta^{(2)}(\qt{1,2})$ with a
larger weight for smaller $a$.

If we integrate the above $\energy_{(0)}(k_\perp)$ over $\kt$ the
initial energy density becomes $a$-independent and should be reduced
to Eq.~(\ref{eq:energy-confirm}).  As we will discuss later, the
spectrum should follow the perturbative time evolution, and then we
will have the energy density as a function of time that is
$a$-dependent and UV finite (but $m$ dependent).

%%%   Second Order   %%%

\subsection{Second Order}

We are now going into the contributions of $\mathcal{O}(\tau^2)$.
There arise two kinds of terms with and without spatial derivatives.
The derivative terms bring about the UV singularity from the Gaussian
average in the MV model.  The non-derivative terms have more gauge
fields in place of derivatives.  We will discuss each contribution in
order.

%--   Longitudinal fields   --%

\subsubsection{Longitudinal fields}

We calculate first the longitudinal field energy of
$\mathcal{O}(\tau^2)$.  Because the longitudinal field has an
$\mathcal{O}(\tau^0)$ contribution, the second-order terms in the
energy come from the cross terms between the zeroth and the second
order in the fields.  That is,
\begin{align}
 \energy^L_{(2)} &= \Bigl\langle \tr\bigl[ 2E^\eta_{(0)}E^\eta_{(2)}
  +2B^\eta_{(0)}B^\eta_{(2)} \bigr]\Bigr\rangle \notag\\
 &= \frac{1}{2}\Bigl\langle \tr\bigl[ E^\eta_{(0)}D_{j(0)}D_{j(0)}
  E^\eta_{(0)} + B^\eta_{(0)}D_{j(0)}D_{j(0)}B^\eta_{(0)}
  \bigr]\Bigr\rangle \,.
\end{align}
The derivative part yields
\begin{align}
 &\frac{1}{2}\Bigl\langle \tr\bigl[ E^\eta_{(0)}\del^2 E^\eta_{(0)}
  + B^\eta_{(0)}\del^2 B^\eta_{(0)} \bigr]\Bigr\rangle \notag\\
 &= 2g^2\Nc(\Nc^2-1)\langle\alpha\alpha\rangle
  \langle\alpha\del^2\alpha\rangle \notag\\
 &= 6g^6\mu_A^4 \del^2L(\zero)\Bigl[ 2(\del^2)^2L(\zero)
  -3g^4\mu_A^2\bigl(\del^2L(\zero)\bigr)^2
  \Bigr] \notag\\
 &= -g^6\mu_A^4\cdot \frac{3}{2\pi^2}\Lambda^2
  \ln\frac{\Lambda}{m} + g^{10}\mu_A^6\cdot
  \frac{9}{4\pi^3}\biggl[\ln\frac{\Lambda}{m}\biggr]^3 \,.
\end{align}
The non-derivative part, on the other hand, gives
\begin{align}
 &-\frac{1}{2}g^2\Bigl\langle\tr\Bigl[ E^\eta_{(0)}\bigl[A_{j(0)},[A_{j(0)},
  E^\eta_{(0)}]\bigr] \notag\\
 &\qquad\qquad
  + B^\eta_{(0)}\bigl[A_{j(0)},[A_{j(0)},B^\eta_{(0)}]\bigr]\Bigr]
  \Bigr\rangle \notag\\
 &= -7g^4\Nc^2(\Nc^2\!-\!1)\langle\alpha\alpha\rangle^3
  = -g^{10}\mu_A^6 \frac{7\cdot9}{8\pi^3} \biggl[\ln
  \frac{\Lambda}{m}\biggr]^3 \,.
\end{align}
The sum of these gives rise to the second-order term of the
longitudinal field energy: 
\begin{equation}
 \energy^L_{(2)} = -g^6\mu_A^4\cdot \frac{3}{2\pi^2}\Lambda^2
  \ln\frac{\Lambda}{m}
  - g^{10}\mu_A^6\cdot \frac{45}{8\pi^3}
  \biggl[\ln\frac{\Lambda}{m}\biggr]^3 \,.
\label{eq:EL_exp}
\end{equation}

%--   Transverse fields   --%

\subsubsection{Transverse fields}

The computation for the transverse fields is parallel to the previous
case.  Since the Hamiltonian~(\ref{eq:hamiltonian}) possesses
$1/\tau^2$ in front of the transverse fields, the square of the
$\mathcal{O}(\tau^2)$ fields contributes to the energy at
$\mathcal{O}(\tau^2)$.  Note that there is no $\mathcal{O}(\tau^0)$
transverse field.  We thus evaluate,
\begin{align}
 \energy^T_{(2)} &= \Bigl\langle \tr\bigl[ E^i_{(2)}E^i_{(2)}
  + B^i_{(2)}B^i_{(2)} \bigr] \Bigr\rangle \notag\\
 &= \frac{1}{4}\Bigl\langle \tr\bigl[ D_{j(0)}E^\eta_{(0)}
  D_{j(0)}E^\eta_{(0)} + D_{j(0)}B^\eta_{(0)}D_{j(0)}B^\eta_{(0)}
  \bigr] \Bigr\rangle \,.
\end{align}
The derivative terms give
\begin{align}
 &\frac{1}{4}\Bigl\langle \tr\bigl[ \partial_j E^\eta_{(0)}
  \partial_j E^\eta_{(0)} + \partial_j B^\eta_{(0)}\partial_j B^\eta_{(0)}
  \bigr] \Bigr\rangle \notag\\
 &= g^2\Nc(\Nc^2-1)\langle\alpha\alpha\rangle
  \langle\partial_j\alpha\partial_j\alpha\rangle \notag\\
 &= -3g^6\mu_A^4 \del^2L(\zero)\Bigl[ 2(\del^2)^2L(\zero)
  -3g^4\mu_A^2\bigl(\del^2L(\zero)\bigr)^2\Bigr] \notag\\
 &= g^6\mu_A^4\cdot \frac{3}{4\pi^2}\Lambda^2
  \ln\frac{\Lambda}{m} - g^{10}\mu_A^6\cdot
  \frac{9}{8\pi^3}\biggl[\ln\frac{\Lambda}{m}\biggr]^3 \,,
\end{align}
which is nothing but a negative half of the corresponding derivative
part in the longitudinal fields.  We can also check that the
non-derivative terms are a negative half of the longitudinal
counterparts as well.  Hence we find
\begin{equation}
 \energy^T_{(2)} = -\frac{1}{2}\energy^L_{(2)} \,.
\label{eq:transverse}
\end{equation}

%%%%%%%%%%   TIME EVOLUTION -- Step III)   %%%%%%%%%%

\section{TIME EVOLUTION -- Step III)}
\label{sec:evolution}

We must take the limit of $\Lambda\to\infty$ in the end because the UV
singularity is not physical in contrast to the IR property which is
physical.  In the situation with large $\Lambda$ only the derivative
terms $\propto\Lambda^2$ are predominant in the $\mathcal{O}(\tau^2)$
corrections.  We shall therefore drop the non-derivative contributions
for the moment.  In fact we could take account of those effects and
find them negligible in the early-time region.

%%%   ENERGY DENSITY   %%%

\subsection{Energy Density}

In Ref.~\cite{Fukushima:2007ja} one of the authors proposed the
following logarithmic ansatz to resum the UV diverging terms:
\begin{align}
 \energy(\tau) &\simeq \energy_{(0)} + \energy_{(2)}\tau^2 \notag\\
 &\simeq g^6\mu_A^4\cdot\frac{3}{4\pi^2}\ln\frac{\Lambda^2}{m^2}
  \biggl(\ln\frac{\Lambda^2}{m^2} -\frac{1}{2}\Lambda^2\tau^2
  \biggr) \notag\\
 &\to g^6\mu_A^4\cdot \frac{3}{4\pi^2} \biggl(
  \ln\biggl[\frac{\Lambda^2}{m^2 + \quart m^2\Lambda^2\tau^2}
  \biggr]\biggr)^2 \,.
\label{eq:log-ansatz}
\end{align}
A nice feature of this ansatz is that the $\Lambda\to\infty$ limit
is well defined, i.e.,
\begin{equation}
 \energy(\tau)=\frac{1}{g^2}(g^2\mu_A)^4 \frac{3}{\pi^2} \biggl[
  \ln\frac{2}{m\tau} \biggr]^2 \,.
\label{eq:simple}
\end{equation}
This is a concise pocket formula.  We will later confirm that
Eq.~(\ref{eq:simple}) works well to give a nice approximation
to the perturbative time evolution of the Bessel function.

In the rest of this section let us elaborate the correct resummation
of the the highest-order divergent terms at each order of $\tau$. The
idea is simple:  We have understood that the UV singularity is
attributed to the spatial derivative.  Then, it is natural to
anticipate that the solution of the equations of motion with the most
singular spatial derivatives retained resums the UV singularity. This
is indeed the case, as is explicitly done here.

The highest-derivative contribution is the solution of the equations
of motion:
\begin{align}
 \partial_\tau E^i &= \tau\partial_j\bigl(\partial_j A_i
  -\partial_i A_j\bigr) = \tau\del^2 P^T_{ij}A_j \,,
\label{eq:pert_i}\\
 \partial_\tau E^\eta &= \frac{1}{\tau}\del^2 A_\eta \,.
\label{eq:pert_eta}
\end{align}
From the above differential equations for $A_i$ and $A_\eta$, the time
dependence is deduced as
\begin{equation}
 \begin{split}
 A_i(\tau,\kt) &= A_{i(0)} J_0(k_\perp\tau) \,,\\
 A_\eta(\tau,\kt) &= A_{\eta(2)} \frac{2\tau}{k_\perp} J_1(k_\perp\tau) \,.
 \end{split}
\end{equation}
We have dropped the transverse projection $P^T_{ij}$ because the
initial condition $E^i_{(2)}$ given as Eq.~(\ref{eq:e_second}) is
transverse in the leading order of the spatial derivative.  We can
find $E^i$ and $E^\eta$ with the use of the definition of the
canonical momenta~(\ref{eq:mom-i}) and (\ref{eq:mom-eta}).  In the
leading order of the spatial derivative we have
$B^\eta=\partial_1A_2-\partial_2A_1$ and
$B^i=\epsilon^{ij}\partial_jA_\eta$, so that $B^\eta$ and $B^i$ have
the time-dependence same as $A_i$ and $A_\eta$, respectively.  In
summary, we have
\begin{align}
 \begin{split}
 E^\eta(\tau,\kt) &= E^\eta_{(0)}(\kt)J_0(k_\perp\tau) \,,\\
 E^i(\tau,\kt) &= E^i_{(2)}(\kt)\frac{2\tau}{k_\perp}J_1(k_\perp\tau) \,,\\
 B^\eta(\tau,\kt) &= B^\eta_{(0)}(\kt)J_0(k_\perp\tau) \,,\\
 B^i(\tau,\kt) &= B^i_{(2)}(\kt)\frac{2\tau}{k_\perp}J_1(k_\perp\tau) \,.
 \end{split}
\end{align}
It follows that the longitudinal field energy component can be summed
up to be
\begin{equation}
 \energy^L(\tau,\kt) = \frac{6}{\pi m^2}g^6\mu_A^4 \,
  \mathcal{T}(a ; k_\perp/m)\bigl[J_0(k_\perp\tau)\bigr]^2
\end{equation}
with $\Nc=3$ substituted.  It is easy to see that the
$\mathcal{O}(\tau^2)$ term exactly corresponds to the $\Lambda^2$
terms in $\energy^L_{(2)}$ if expanded.  Likewise, the transverse
field energy becomes
\begin{equation}
 \energy^T(\tau,\kt) = \frac{6}{\pi m^2}g^6\mu_A^4 \,
  \mathcal{T}(a ; k_\perp/m)\bigl[J_1(k_\perp\tau)\bigr]^2 \,,
\end{equation}
whose coefficient is fixed so as to reproduce
Eq.~(\ref{eq:transverse}) in the $\tau$-expansion.  The total energy
density at each momentum thus evolves as
\begin{equation}
 \energy(\tau,\kt) = \energy_{(0)}(\kt)
  \Bigl(\bigl[J_0(k_\perp\tau)\bigr]^2+\bigl[J_1(k_\perp\tau)\bigr]^2
  \Bigr) \,,
\end{equation}
and the integrated one is
\begin{align}
 \energy(\tau) &= \frac{1}{2\pi} \int dk_\perp k_\perp\,
  \energy(\tau,\kt) \notag\\
 &= \frac{3}{\pi^2}\cdot\frac{1}{g^2}(g^2\mu_A)^4 \,
  I_E(a ; m\tau) \,,
\label{eq:initial-energy}
\end{align}
where we have defined a dimensionless function,
\begin{equation}
 I_E(a ; m\tau) = \int_0^\infty d\xi\,\xi\,\mathcal{T}(a ; \xi)
  \Bigl\{\bigl[J_0(\xi m\tau)\bigr]^2
  + \bigl[J_1(\xi m\tau)\bigr]^2 \Bigr\} \,.
\label{eq:IE}
\end{equation}
Although our derivation is rather straightforward, this time
dependence according to the Bessel function is exactly the same as the
one given in Ref.~\cite{Kovchegov:2005ss}.  It should be noted here
that this Bessel function form has been discussed in the earliest work
in Ref.~\cite{Kovner:1995ja} and confirmed in the numerical simulation
in Ref.~\cite{Krasnitz:1998ns} in the \textit{weak field} limit.
However, we do not assume the weak field limit at all in our argument
but the UV dominance at $\tau\ll1/Q_s$ is sufficient to justify this
time dependence.  The same evolution is also found in the Abelian
model in Ref.~\cite{Fujii:2008dd}.

\begin{figure}
 \includegraphics[width=8cm]{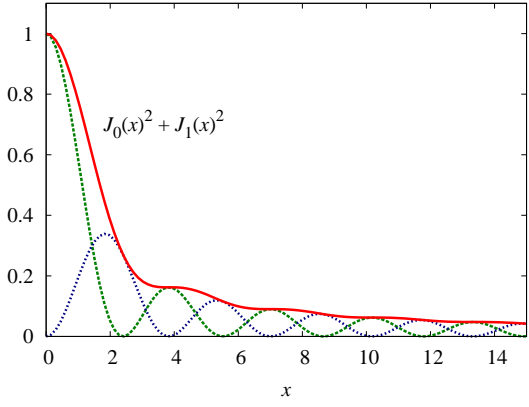}
 \caption{$[J_0(x)]^2+[J_1(x)]^2$ as a function of $x$ which goes to
   zero as $2/(\pi x)$ at large $x$.  The dotted curve starting from
   $1$ represents $[J_0(x)]^2$ and the other one starting from $0$
   represents $[J_1(x)]^2$.}
 \label{fig:bessel}
\end{figure}

Our expression for the initial energy
density~(\ref{eq:initial-energy}) is a function of the proper time
$\tau$, the QCD coupling $g$, the MV model scale $g^2\mu_A$, and the
IR cutoff $m$ or its ratio $a$ to $g^2\mu_A$.  The Bessel function
$[J_0(k_\perp\tau)]^2+[J_1(k_\perp\tau)]^2$ encompasses the time
dependence which we plot in Fig.~\ref{fig:bessel}.  Interestingly
enough, the sum of two is a smooth function, though each of
$[J_0(k_\perp\tau)]^2$ and $[J_1(k_\perp\tau)]^2$ is an oscillatory
function of $k_\perp\tau$.  This is reminiscent of the trigonometric
function.  Moreover, we can prove that
\begin{equation}
 \bigl[ J_0(k_\perp\tau) \bigr]^2 + \bigl[ J_1(k_\perp\tau) \bigr]^2
  \to \frac{2}{\pi k_\perp\tau} \,,
\end{equation}
asymptotically in the $k_\perp\tau\to\infty$ limit.  This analytical
feature results in the behavior of
$\energy(\tau\to\infty)\propto 1/\tau$ that is the same as the
free-streaming expansion~\cite{Kovchegov:2005ss}.

%%%   Gluon Distribution   %%%

\subsection{Gluon Distribution}

As for the gluon distribution, we shall adopt here a working
definition in accord with the preceding
works~\cite{Kovner:1995ja,Gyulassy:1997vt,Krasnitz:1998ns,%
Krasnitz:2001qu,Lappi:2003bi,Dumitru:2001ux}.  That is, we assume a
harmonic oscillator and count the number of quanta.  The concrete
procedure is that we divide the field energy with momentum $\kt$ by
the (free) gluon energy quanta $k_\perp$ to infer the gluon
distribution contained in the fields, which is consistent with the
multiplicity computation by the reduction formula~\cite{Gelis:2006yv}.
We define in this way the gluon distribution as
\begin{equation}
 n(\tau,\kt) = \frac{1}{\sqrt{k_\perp^2+m^2}}\,\energy(\tau,\kt) \,,
\end{equation}
where we put the gluon mass $m$.  From this we readily reach the
expression for the gluon distribution as
\begin{align}
 n(\tau) &= \frac{1}{2\pi} \int dk_\perp k_\perp\, n(\tau,\kt)
  \notag\\
 &= \frac{3}{\pi^2 m}\cdot\frac{1}{g^2}(g^2\mu_A)^4 \,
  I_N(a ; m\tau) \,,
\label{eq:initial-dist}
\end{align}
where
\begin{equation}
 I_N(a ; m\tau) = \int_0^\infty d\xi\,\frac{\xi\,\mathcal{T}(\xi)}
  {\sqrt{\xi^2+1}} \Bigl\{\bigl[J_0(\xi m\tau)\bigr]^2
  + \bigl[J_1(\xi m\tau)\bigr]^2 \Bigr\} \,.
\label{eq:IN}
\end{equation}
It should be mentioned that $dn/d^3 k$ at $\tau=0$ goes like
$\propto (1/k_\perp^4)\ln(k_\perp/m)$ for large $\kt$ which reproduces
the perturbative results known as the Gunion-Bertsch
multiplicity~\cite{Kovner:1995ja,Gunion:1981qs}.  To compare with the
observed multiplicity, we have to assume the parton-hadron duality and
the entropy conservation to interpret the initial gluon distribution
as converted to the particle multiplicity.

%%%   Numerics   %%%

\subsection{Numerics}

\begin{figure}
 \includegraphics[width=8cm]{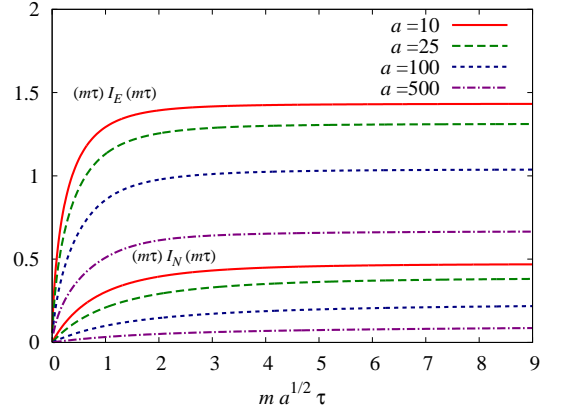}
 \caption{$(m\tau)I_E(m\tau)$ and $(m\tau)I_N(m\tau)$ used in
   Eqs.~(\ref{eq:xIE}) and (\ref{eq:xIN}), respectively, as a function
   of $m\sqrt{a}\,\tau$ for various $a$.}
 \label{fig:I}
\end{figure}

We are now ready to give a numerical estimate for the initial energy
and the multiplicity from our formulation.  The key equations are
Eqs.~(\ref{eq:initial-energy}), (\ref{eq:IE}),
(\ref{eq:initial-dist}), and (\ref{eq:IN}).  From these expressions we
can write the total energy and the multiplicity per rapidity as
\begin{align}
 \frac{dE(\tau)}{d\eta} &= \pi R_A^2\, \tau\energy(\tau) \notag\\
  &= \frac{3\pi R_A^2}{\pi^2 m}\cdot\frac{1}{g^2}(g^2\mu_A)^4
  (m\tau)I_E(a ; m\tau) \,,
\label{eq:xIE}\\
 \frac{dN(\tau)}{d\eta} &= \pi R_A^2\, \tau n(\tau) \notag\\
  &= \frac{3\pi R_A^2}{\pi^2 m^2}\cdot\frac{1}{g^2}(g^2\mu_A)^4
  (m\tau)I_N(a ; m\tau) \,.
\label{eq:xIN}
\end{align}
We plot $(m\tau)I_E(a ; m\tau)$ and $(m\tau)I_N(a ; m\tau)$
appearing in the above expressions in Fig.~\ref{fig:I} for various
$a$ using $\mathcal{T}(a ; k_\perp/m)$ shown in Fig.~\ref{fig:integ}.
It is clear that both $(m\tau)I_E(m\tau)$ and $(m\tau)I_N(m\tau)$
behave approaching constant for large $m\tau$ as we have mentioned
before.

\begin{figure}
 \includegraphics[width=8cm]{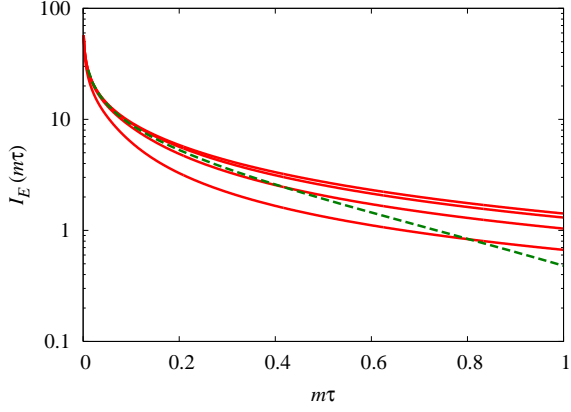}
 \caption{Comparison between the log-ansatz $(\ln[2/m\tau])^2$ (green
   dashed curve) and the numerically evaluated $I_E(m\tau)$ (red solid
   curves) for $a=10$, $25$, $100$, and $500$ from the top to the
   bottom.}
 \label{fig:compare}
\end{figure}

The validity of the CGC (classical field) approximation breaks down
when the system becomes dilute near
$g^2\mu_A\tau\sim m\sqrt{a}\,\tau \sim 1$.  In the very early stage of
the evolution, on the other hand, the UV modes dominate in the
physical quantities and the logarithmic ansatz~(\ref{eq:log-ansatz})
works quite well.  We see this fact from the comparison in
Fig.~\ref{fig:compare} which is the same plot as Fig.~\ref{fig:I}
without the factor $m\tau$.  This is quite a non-trivial statement.  We
started from $\tau=0$ with the exact initial condition provided by the
MV model.  Then we found that, as long as $m\sqrt{a}\,\tau$ is much
smaller than unity (or simply $m\tau<1$), only the UV modes control
the time evolution of the energy and there is only little effect from
the IR sector which changes with $a$.  Hence, the remaining $a$
dependence is only the trivial scaling in terms of $m$ with
$m\sqrt{a}=\sqrt{\Nc/2}\,g^2\mu_A$ fixed as long as $\tau$ is small.

From Fig.~\ref{fig:I} we can give physical numbers for the energy and
the multiplicity deduced in the $1/\tau$-scaling region.  We first set
$\pi R_A^2\approx150\fm^2$ and $g\approx 2$ which are less ambiguous.

In the case of $a=10$, we have obtained $m\simeq 0.64Q_s$ and
$(g^2\mu_A)^2\simeq 2.70Q_s^2$.  The asymptotic values of
$(m\tau)I_E(a,m\tau)$ and $(m\tau)I_N(a,m\tau)$ at $\tau\to\infty$ 
are respectively $1.43$ and $0.47$.
Thus, we have
$dE/d\eta=1.4\times 10^4\GeV$ and $dN/d\eta=5.0\times 10^3$ in the
case of $Q_s^2=2\GeV^2$.  If $Q_s^2=1\GeV^2$, $dE/d\eta$ and $dN/d\eta$
get smaller by factors $2^{3/2}$ and 2, respectively.
For other values of $a$, we can get  the estimates in the same way. We
summarize our results in Table~\ref{tab:e-n}.  We note that $c$ listed
in Table~\ref{tab:e-n} is the gluon liberation factor defined
by~\cite{Mueller:1999fp,Kovchegov:2000hz}
\begin{equation}
 \frac{1}{\pi R_A^2}\frac{dN}{d\eta}
 = c\,\frac{\Nc^2-1}{\pi g^2\Nc}\,Q_s^2 \,.
\end{equation}

\begin{table}
 \begin{tabular}{|c||c|c|c|}
\hline
 $a$  &  $dE/d\eta$ (GeV)  &  $dN/d\eta$  &  $c$\\
\hline
 $\quad 10\quad$ & ~~$(4.9 - 14)\times 10^3$~~ &
 ~~$(2.5 - 5.0)\times 10^3$~~ & ~~$3.04$~~\\
\hline
 $25$ & $ (4.9 - 14)\times 10^3$ & $(4.0 - 7.9)\times 10^3$ & $4.86$\\
\hline
 $100$ & $(3.5 - 10)\times 10^3$ & $(5.6 - 11)\times 10^3$ & $6.88$\\
\hline
 $500$ & $(2.6 - 7.3)\times 10^3$ & $(7.7 - 16)\times 10^3$ & $9.47$\\
\hline
 \end{tabular}
 \caption{Total energy and gluon distribution estimated in our
   formulation for various $a$ with the choice of
   $Q_s^2=1$--$2\GeV^2$.  [Lower and upper values correspond to
   $Q_s^2=1\GeV^2$ and $Q_s^2=2\GeV^2$, respectively.]  We also list
   the gluon liberation factor $c$.}
 \label{tab:e-n}
\end{table}

It is interesting to note that the energy is less sensitive to $a$ in
the range of our interest while the multiplicity rises significantly
with increasing $a$.  This is because $a$ controls the IR scale $m$ to
which the average gluon energy $(dE/d\eta)/(dN/d\eta)$ is
proportional.  In other words, with increasing $a$ we include more IR
modes, which contribute to $dN/d\eta$ more than $dE/d\eta$.

%%%%%%%%%%   IMPROVEMENT -- Step IV)   %%%%%%%%%%

\section{IMPROVEMENT -- Step IV)}
\label{sec:improve}

Let us consider here a possible improvement of our estimate.  The UV
dominance cannot last at later time, and the nonlinear effect may
become manifest in the intermediate stage of the evolution.  We then
need include less singular terms from the nonlinear interactions
appearing in Eq.~(\ref{eq:EL_exp}).  In order to estimate the size of
this nonlinear effect we take the following strategy:  In the Gaussian
approximation we adopt here,  the nonlinear terms in the Yang-Mills
equations of motion produce the mean-field contributions like the mass
term, that would modify Eqs.~(\ref{eq:pert_i}) and (\ref{eq:pert_eta})
as
\begin{align}
 \partial_\tau E^{Ti}
&= \tau\bigl(\del^2 -\kappa_{_T}g^2\Nc\langle\alpha\alpha\rangle
  \bigr)A_i^T \,,
\label{eq:ET} \\ 
 \partial_\tau E^\eta
&= \frac{1}{\tau}\bigl( \del^2 -\kappa_{\eta}g^2\Nc
  \langle\alpha\alpha\rangle \bigr) A_\eta \,,  
\label{eq:Eeta}
\end{align}
where $\kappa_{_T}$ and $\kappa_{\eta}$ are the coefficients coming from
contractions of the gauge fields;
$A_i^a A_j^b A_\eta^c\to \delta^{ab}\delta_{ij}
\langle\alpha\alpha\rangle A_\eta^c$ and
$A_i^a A_j^b A_k^{Tc}\to \delta^{ab}\delta_{ij}
\langle\alpha\alpha\rangle A_k^{Tc}$, etc.  Noting that the initial
$A_i$ has no correlation with $A_i^T$ in the leading order of the
derivative, on the one hand, we find $\kappa_{_T}=\kappa_{\eta}=2$ by
these replacements in the equations of motion, though this might
underestimate the effect.  On the other hand, if we determine
$\kappa_{_T}$ and $\kappa_{\eta}$ to reproduce the expanded energy in
Eqs.~(\ref{eq:EL_exp}) and (\ref{eq:transverse}) up to
$\mathcal{O}(\tau^{2})$, we conclude $\kappa_{_T}=\kappa_{\eta}=7$,
which we shall adopt here.

In Eqs.~(\ref{eq:ET}) and (\ref{eq:Eeta}), we have estimated
$\langle A_i A_i\rangle$ at $\tau=0$ and we did not include
$\langle A_\eta A_\eta\rangle$ which  is $\mathcal{O}(\tau^{4})$.
From Eqs.~(\ref{eq:corr}) and (\ref{eq:log}) we know that
$g^2\Nc\langle\alpha\alpha\rangle=(\Nc/4\pi)(g^2\mu_A)^2\ln[\Lambda/m]
=(m^2 a)/(2\pi)\ln[\Lambda/m]$ with $a$ introduced in
Eq.~(\ref{eq:alpha}).  Our discussion on the log-ansatz suggests that
this $\ln[\Lambda/m]$ is taken over by $\ln[2/m\tau]$ at finite $\tau$
(see Fig.~\ref{fig:compare}), and it takes the value around
$\ln 20\simeq 3$ to $\ln 200\simeq 5.3$ for $m\tau=0.01\sim0.1$ of our
interest (so that $m\sqrt{a}\,\tau$ is below unity).  During this
interval, therefore, we may replace this logarithmic function by a
constant of order $\mathcal{O}(1)\sim\mathcal{O}(10)$, in effect.  Let
us assume rather arbitrarily the value $\pi$ for $\ln[\Lambda/m]$,
which is within the range from 3 to 5.3, so that the mean-field effect
appears only through $k_\perp^2\to k_\perp^2+3.5m^2 a$ in the argument
of the Bessel function.

\begin{figure}
 \includegraphics[width=8cm]{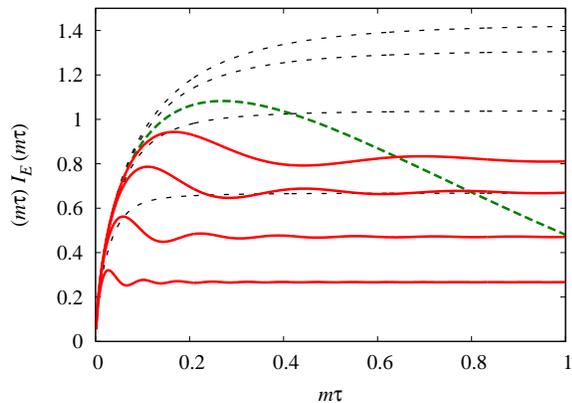}
 \caption{Comparison between the log-ansatz $(m\tau)(\ln[2/m\tau])^2$
   (green dashed curve) and the numerically evaluated
   $(m\tau)I_E(a;m\tau)$ with (red solid curves) and without (black
   dotted curves) $k_\perp^2\to k_\perp^2+3.5m^2 a$ for $a=10$, $25$,
   $100$, and $500$ from the top to the bottom.}
 \label{fig:compare_m}
\end{figure}

Figure~\ref{fig:compare_m} shows how $(m\tau)I_E(a;m\tau)$ changes
due to the replacement of $k_\perp^2\to k_\perp^2+3.5m^2 a$.  We
summarize our final results with nonlinear terms in
Tab.~\ref{tab:e-n2}.  It turns out that the nonlinear effect leads to
an interesting feature that the dependence of $dN/d\eta$ on changing
$a$ is mild, while $dE/d\eta$ retains almost the same (or stronger)
decreasing behavior as it is previously in Tab.~\ref{tab:e-n}.

\begin{table}
 \begin{tabular}{|c||c|c|c|}
\hline
 $a$  &  $dE/d\eta$ (GeV)  &  $dN/d\eta$  &  $c$\\
\hline
 $\quad 10\quad$ & ~~$(2.8 - 7.8)\times 10^3$~~ &
 ~~$(0.7 - 1.4)\times 10^3$~~ & ~~$0.86$~~\\
\hline
 $25$ & $(2.5 - 7.0)\times 10^3$ & $(0.9 - 1.8)\times 10^3$ & $1.10$\\
\hline
 $100$ & $(1.6 - 4.5)\times 10^3$ & $(1.0 - 1.9)\times 10^3$ & $1.18$\\
\hline
 $500$ & $(1.0 - 2.9)\times 10^3$ & $(1.0 - 1.9)\times 10^3$ & $1.17$\\
\hline
 \end{tabular}
 \caption{Total energy and gluon distribution estimated in our
   formulation with nonlinear terms for various $a$ with the
   choice of $Q_s^2=1$--$2\GeV^2$.}
 \label{tab:e-n2}
\end{table}

%%%%%%%%%%   DISCUSSIONS AND OUTLOOKS   %%%%%%%%%%

\section{DISCUSSIONS AND OUTLOOKS}
\label{sec:discussions}

We have shown by explicit calculations in the MV model that the UV
components are dominant in the energy density at $\tau=0$ produced by
the collisions.  From this observation we have evolved the exact
initial energy provided by the MV model via the free field equation.
We did not require the assumption of small field amplitude, but the UV
dominance allows us to drop nonlinear terms in the early time
evolution.

To extend the validity region of the estimate, we have augmented the
time evolution with the effect of nonlinear terms in the Gaussian-type
approximation.  Assuming that the IR cutoff $m$ arises from the
quantum evolution $m\simeq g\mu_A$ or from non-perturbative dynamics
$m\simeq \LQCD$, we have the IR parameter $a=10$--$40$.  In this range
of $a$, we have found that the nonlinear terms tend to reduce the
$a$-dependence of the physical observables.  Let us then take $a=25$.
Regarding the relevant saturation scale, we take $Q_s^2=1\GeV^2$ here
because the discussion in Ref.~\cite{Lappi:2008eq} suggests that
$Q_s^2$ is close to $1\GeV^2$ than $2\GeV^2$.  Thus, what we get for a
realistic situation at RHIC is $dE/d\eta\simeq 2.5\times10^3\GeV$ and
$dN/d\eta\simeq 0.9\times10^3$ with $c=1.10$.  Our result favors a
larger value for $c$ than the early numerical
simulations~\cite{Krasnitz:1998ns}, but is consistent with the
analytical calculation
$c\approx 2\ln2\approx 1.4$~\cite{Kovchegov:2000hz} and agrees well
with the latest numerical
evaluation~\cite{Lappi:2007ku,Lappi:2008eq}.

Our calculation seems to underestimate the total multiplicity whose
empirical value is $\simeq 1150$ at RHIC.\ \ We should, however, be
aware that $dN/d\eta$ here signifies the total number of gluons
produced initially.  Therefore, though $dN/d\eta$ should not be far
from the observed multiplicity, one should not expect a perfect
quantitative agreement.  In this sense the level of the agreement we
obtained is quite acceptable and we would say that our results are
consistent with the empirical values.

A rough estimate of the average energy per gluon is
$(dE/d\eta)/(dN/d\eta)\simeq 2.8\GeV$.  This value is greater than the
largest energy anticipated in Ref.~\cite{Lappi:2003bi}, above which
the CGC description might not make sense.  This speculation is not
quite correct, however.  As is known and also clear in
Fig.~\ref{fig:integ}, the MV model is smoothly connected to the
leading-order perturbative calculation with increasing $\kt$, so that
the spectrum has a perturbative tail in the high momentum region and
the gluon energy is not bounded by the saturation scale.  In contrast,
the nonlinear effect of the CGC will be the most important for the
gluons with momentum of order of $Q_s$, and the average energy per
gluon scales with $Q_s$, which is indeed the case in our formulae.

It is also intriguing to mention on the ``formation time'', which can
roughly read from Fig.~\ref{fig:compare_m} for the $a=25$ case.  We
see that the curve gets flattened around $m\tau\simeq 0.2$.  Since
$m=0.36Q_s$, the ``formation time'' denoted as $\tau_{D}$ which
controls the convergence of $\tau\energy(\tau)$ (see discussions in
the second paper of Ref.~\cite{Krasnitz:1998ns}) is estimated as
$\tau_{D}\simeq 0.56/Q_s\simeq 0.1\fm$ or $g^2\mu_A\tau\simeq 0.8$.
This is pretty short as compared with the expected thermalization time
$\lesssim 1\fm$.  This quantitative estimate of the ``formation time''
might deserve further investigation in the context of the early
thermalization problem.

The future applications, in addition to the early thermalization
problem, cover various aspects of phenomenology as follows.  1) The
rapidity dependence with changing $\mu_A$ by $Q_s(x)$ gives the whole
multiplicity distribution.  2) Finite size $\rho^{(1,2)}(\xt)$ in the
transverse plane yields the initial eccentricity at finite impact
parameter.  3) The energy-momentum tensor specifies the initial
condition connected to the kinematic or hydrodynamic description.  4)
The (augmented) UV time evolution of the background fields is useful
to attack the rapidity dependent instability on top of
them~\cite{Romatschke:2005pm}.  What is addressed here is the starting
point for all of these issues.  We note that the point 2) is
especially important because the initial eccentricity obtained in the
KLN model~\cite{Kharzeev:2004if,Kharzeev:2002pc} is prevailing as the
CGC estimate at present~\cite{Hirano:2004en}, though the confirmation
from the MV model is definitely necessary~\cite{Lappi:2006xc}.

Finally, we close our discussions with a conclusion that our
analytical estimates turn out to be consistent with the numerical
simulations in spite of the difference in their model
definitions~\cite{Fukushima:2007ki}.  This is an important check for
the MV model foundation.

\acknowledgments
The authors thank Raju Venugopalan for discussions.  
The work of H.~F.\ is supported in part by Grants-in-Aid 
(19540273, 19540269) of MEXT.\ \
K.~F.\ is supported by Japanese MEXT grant No.\ 20740134 and also
supported in part by Yukawa International Program for Quark Hadron
Sciences (YIPQS).
Y.~H.\ is supported in part by the RIKEN BNL Research 
Center and by the U.S. Department of Energy under
Cooperative Research Agreement No.\ DE-AC02-98CH10886.

\end{document}